\PassOptionsToPackage{table,xcdraw}{xcolor}
\documentclass[sigconf]{acmart}

\setcopyright{acmcopyright}
\copyrightyear{2020}
\acmYear{2020}

\acmConference[ESEC/FSE 2020]{The 28th ACM Joint European Software Engineering Conference and Symposium on the Foundations of Software Engineering}{8--13 November, 2020}{Sacramento, California, United States}

\usepackage{framed}
\usepackage{paralist}
\usepackage{xspace}
\newcommand{\app}{\emph{LogCleaner}\xspace}
\usepackage[binary-units]{siunitx}
\usepackage[linesnumbered, ruled]{algorithm2e}

\usepackage{todonotes}

\usepackage{multirow}
\definecolor{lightgray}{rgb}{0.83, 0.83, 0.83}

\usepackage{tikz}

\begin{document}

\title[Effective Removal of Operational Log Messages: an Application to Model Inference]{Effective Removal of Operational Log Messages:\\ an Application to Model Inference}

\author{Donghwan Shin}
\affiliation{%
  \institution{University of Luxembourg}
  \country{Luxembourg}
}
\email{donghwan.shin@uni.lu}

\author{Domenico Bianculli}
\orcid{0000-0002-4854-685X}
\affiliation{%
  \institution{University of Luxembourg}
  \country{Luxembourg}
}
\email{domenico.bianculli@uni.lu}

\author{Lionel Briand}
\affiliation{%
  \institution{University of Luxembourg}
  \country{Luxembourg}
}
\email{lionel.briand@uni.lu}

\begin{abstract}
Model inference aims to extract accurate models from the execution logs of 
software systems. However, in reality, logs may contain some ``noise'' that 
could deteriorate the performance of model inference. One form of noise can 
commonly be found in system logs that contain not only \emph{transactional} 
messages---logging the functional behavior of the system---but also 
\emph{operational} messages---recording the operational state of the system 
(e.g., a periodic heartbeat to keep track of the memory usage). In low-quality 
logs, transactional and operational messages are randomly interleaved, leading 
to the erroneous inclusion of operational behaviors into a system model, that 
ideally should only reflect the functional behavior of the system. It is 
therefore important to remove operational messages in the logs before 
inferring models.

In this paper, we propose \app, a novel technique for  
removing operational logs messages. \app first performs a \emph{periodicity 
analysis} to filter out periodic messages, and then it performs a 
\emph{dependency analysis} to calculate the degree of dependency for all log 
messages and to remove operational messages based on their dependencies. 

The experimental results on two proprietary and 11 publicly available
log datasets show that \app, on average, can accurately remove 98\% of
the operational messages and preserve 81\% of the transactional
messages.  Furthermore, using logs pre-processed with \app decreases
the execution time of model inference (with a speed-up ranging from
1.5 to 946.7 depending on the characteristics of the system) and
significantly improves the accuracy of the inferred models, by increasing their
ability to accept correct system behaviors (+43.8\thinspace $pp$ on average, with
$pp$=percentage points) and to reject
incorrect system behaviors (+15.0\thinspace $pp$ on average).

\end{abstract}




\maketitle

\section{Introduction}\label{sec:intro}

\emph{Model inference} aims to extract models---typically in the form of Finite 
State Machine (FSM)---from the execution logs of software systems. Such behavioral 
models can play a key role in many software engineering tasks, such as program 
comprehension~\cite{Cook:1998:287001}, test case generation~\cite{6200086}, and 
model checking~\cite{clarke2018model}. Over time, a variety of algorithms have been 
proposed to infer 
FSMs~\cite{biermann1972synthesis,Beschastnikh2011Lev,LUO201713} or richer 
variants, such as gFSM (guarded 
FSM)~\cite{walkinshaw2016inferring,mariani2017gk} and gFSM extended with 
transition probabilities~\cite{Emam2018Inf}, to obtain more faithful models.

The accuracy of the models obtained through model inference techniques
depends on the ``quality'' of the input logs: such techniques work on the
assumption that a given log recorded during the execution of a system
faithfully represents the functional behavior of the system. However,
in reality, logs may contain some ``noise'' that, if not properly
removed, could be wrongly captured in the inferred model.  One form
of noise can commonly be found in system logs that contain not only
\emph{transactional} messages---logging the functional behavior of the
system---but also \emph{operational} messages---recording the
operational state of the system (e.g., a periodic heartbeat to keep
track of the memory usage).  In
low-quality, noisy logs, transactional and operational messages are
randomly interleaved, leading to the
inference of inaccurate models. The latter erroneously incorporate
operational behaviors (e.g., a heartbeat) into a system model that
ideally should only reflect the functional behavior of the system.

To improve the effectiveness of model inference techniques (and thus
the accuracy of the inferred models) it is therefore important to
pre-process the logs before inferring models, by identifying and
removing forms of noise, such as randomly interleaved operational
messages.  The manual analysis of logs, to distinguish between
operational and transactional messages, is expensive, cumbersome, and
error-prone.  Furthermore, the identification and removal of
operational messages is even more challenging in contexts where the
system is mainly composed of heterogeneous, 3rd-party components for
which neither the source code nor the documentation are available,
since there is little or no domain knowledge available to drive this process.

Recently, \citet{8030620} proposed an automated technique for
operational message filtering, as part of the pre-processing step for
a log-based anomaly detection technique; the intuition behind this
technique is that randomly interleaved operational messages are more
likely to be independent from others compared to transactional
messages. However, this filtering technique does not systematically
cover various types of dependency between log messages. For example,
if two messages $x$ and $y$ are frequently co-occurred within a small 
distance, the filtering technique deems high the dependency between $x$ and $y$. 
However, it could be the case that $x$ and $y$ are just frequently 
interleaved in random order without any dependency, meaning at least 
one of them is an operational message. Other techniques for
noise filtering in the area of (business) process
mining~\cite{7579568,Zelst2018Filtering,8818438}, mainly target
outliers or infrequent behaviors rather than randomly interleaved
operational messages; therefore they cannot be applied in our context.

In this paper, we propose a novel technique, named \app, for the
identification and removal of operational log messages, which extends
the heuristic originally proposed in reference~\cite{8030620} by
taking into account the dependencies among messages as well as the
intrinsic periodicity of some operational log messages. Specifically,
\app first performs a \emph{periodicity analysis} to filter out
certain messages that are deemed periodic throughout a given set of
logs (e.g., a message that constantly appears every second throughout
the entire log). \app then performs a \emph{dependency analysis} to
calculate the degree of dependency for all log messages and to identify
operational messages based on their dependencies. Since randomly
interleaved operational messages are likely to have smaller degrees
of dependency  than transactional messages, a clustering-based
heuristic can automatically separate operational messages from
transactional messages using this information.

We evaluate the accuracy of \app in removing operational log messages
(and preserving transactional messages) on two proprietary log
datasets from one of our industrial partners in the satellite domain
and 11 publicly available log datasets from the literature.  We also
evaluate how the logs pre-processed through \app affect the cost (in
terms of execution time) and accuracy of a state-of-art model
inference technique (MINT~\cite{walkinshaw2016inferring}). The results
show that \app, on average, can remove 98\% of the operational
messages and preserve 81\% of the transactional messages.
Furthermore, using logs pre-processed with \app decreases the
execution time of model inference (with a speed-up ranging from 1.5 to
946.7 depending on the characteristics of the system) and improves
the accuracy of the inferred models, by increasing their ability to
accept correct system behaviors (+43.8\thinspace $pp$ on average, with
$pp$=percentage points) and  to reject incorrect system
behaviors (+15.0\thinspace $pp$ on average).

To summarize, the main contributions of this paper are:
\begin{itemize}
\item the \app approach for taming the problem of identification and
  removal of operational log messages;
\item a publicly available implementation of \app; and
\item the empirical evaluation of the accuracy of \app and the benefits \app 
brings to an existing model inference technique in terms of execution time and 
accuracy.
\end{itemize}

The rest of the paper is organized as follows. Section~\ref{sec:preliminary} 
gives the basic definitions of logs and a running example that will be used 
throughout the paper. Section~\ref{sec:technique} describes the core algorithms 
of \app, whereas Section~\ref{sec:eval} reports on its evaluation. 
Section~\ref{sec:related-work} discusses related work. 
Section~\ref{sec:conclusion} concludes the paper and provides directions for 
future work.

\section{Preliminaries}\label{sec:preliminary}
A log is a sequence of log entries. A log entry contains a timestamp (recording 
the time at which the logged event occurred) and a log message (with run-time 
information). A log message can be further decomposed~\cite{messaoudi2018search} 
into a fixed part called event template, characterizing the event type, and a variable part, which
contains tokens filled at run time with the values of the event parameters.
For example, the first log entry of the example log $l_\mathit{org}$ shown in Figure~\ref{fig:example-log} 
is composed of the timestamp \texttt{20180625:10:00:01}, 
the template \texttt{\textbf{ping}} $v_1$, 
and the set of parameter values $\{v_1 = \texttt{OK}\}$. Similarly, the second
log entry is composed of the timestamp \texttt{20180625:10:00:01}, the template 
\texttt{\textbf{send}} $v_1$ \texttt{\textbf{via}} $v_2$, and the set of parameter values 
$\{v_1 = \texttt{MSG1}, v_2 = \texttt{CH1}\}$. More formally, let $L$ be the set 
of all logs, $T$ be the set of all event templates, and $P$ be the set of all 
mappings from event parameters to their concrete values. A log $l\in L$ is a 
sequence of log entries $\langle e_1, \dots, e_n \rangle$, with $e_i = (\tau_i, 
t_i, p_i)$, $\tau_i\in \mathbb{N}$, $t_i\in T$, and $p_i\in P$, for $i=1,\dots,n$.
The example log $l_\mathit{org}$ shown in Figure~\ref{fig:example-log} can be 
represented by $l_\mathit{org} = \langle e_1, e_2, \dots, e_{19} \rangle$ where $\tau_1=$ 
\texttt{20180625:10:00:01}, $t_1 =$ \texttt{\textbf{ping}} $v_1$, $p_1 = \{v_1 = 
\texttt{OK}\}$, and so on. In the rest of the paper, we denote a template using 
its first word for simplicity; for example, we say the template 
\texttt{\textbf{ping}} instead of the template \texttt{\textbf{ping}} $v_1$.

\begin{figure}
\begin{tabular}{lll}
\toprule
$l_\mathit{org}$ & $l_\mathit{inter}$ & Log entry (timestamp + message) \\
\midrule
\rowcolor{lightgray} 
$e_1$ & - &  \texttt{20180625:10:00:01 \textbf{ping} OK} \\
$e_2$ & $e_{1'}$ & \texttt{20180625:10:00:01 \textbf{send} MSG1 \textbf{via} CH1} \\
\rowcolor{lightgray} 
$e_3$ & - & \texttt{20180625:10:00:02 \textbf{ping} OK} \\
\rowcolor{lightgray} 
$e_4$ & $e_{2'}$ & \texttt{20180625:10:00:02 \textbf{memory} OK} \\
$e_5$ & $e_{3'}$ & \texttt{20180625:10:00:02 \textbf{check} MSG1} \\
\rowcolor{lightgray} 
$e_6$ & - & \texttt{20180625:10:00:03 \textbf{ping} OK} \\
$e_7$ & $e_{4'}$ & \texttt{20180625:10:00:03 \textbf{check} MSG1} \\
\rowcolor{lightgray} 
$e_8$ & $e_{5'}$ & \texttt{20180625:10:00:03 \textbf{memory} OK} \\
\rowcolor{lightgray} 
$e_9$ & - & \texttt{20180625:10:00:04 \textbf{ping} OK} \\
\rowcolor{lightgray} 
$e_{10}$ & - & \texttt{20180625:10:00:05 \textbf{ping} OK} \\
\rowcolor{lightgray} 
$e_{11}$ & - & \texttt{20180625:10:00:06 \textbf{ping} OK} \\
\rowcolor{lightgray} 
$e_{12}$ & $e_{6'}$ & \texttt{20180625:10:00:06 \textbf{memory} OK} \\
$e_{13}$ & $e_{7'}$ & \texttt{20180625:10:00:06 \textbf{send} MSG2 \textbf{via} CH1} \\
\rowcolor{lightgray} 
$e_{14}$ & - & \texttt{20180625:10:00:07 \textbf{ping} OK} \\
$e_{15}$ & $e_{8'}$ & \texttt{20180625:10:00:07 \textbf{check} MSG2} \\
\rowcolor{lightgray} 
$e_{16}$ & - & \texttt{20180625:10:00:08 \textbf{ping} OK} \\
\rowcolor{lightgray} 
$e_{17}$ & - & \texttt{20180625:10:00:09 \textbf{ping} OK} \\
\rowcolor{lightgray} 
$e_{18}$ & $e_{9'}$ & \texttt{20180625:10:00:09 \textbf{memory} OK} \\
\bottomrule
\end{tabular}
\caption{Logs of the running example (operational messages highlighted
in grey)}
\label{fig:example-log}
\end{figure}

The example log $l_\mathit{org}$ shown in Figure~\ref{fig:example-log} will 
be used as a running example throughout the paper. It has four event templates, 
i.e., $T=\{\texttt{\textbf{send}}, \texttt{\textbf{check}}, 
\texttt{\textbf{ping}}, \texttt{\textbf{memory}}\}$. Among them, 
\texttt{\textbf{send}} and \texttt{\textbf{check}} are transactional
templates that represent the functional behavior of the system;  
\texttt{\textbf{ping}} and \texttt{\textbf{memory}} are operational templates 
that represent the operational state of the system. The operational messages 
are highlighted in grey; as you can see, they are randomly interleaved 
with transactional messages.
Figure~\ref{fig:example-log} also contains a second example log,
$l_\mathit{inter}$, which contains some of the entries of
$l_\mathit{org}$; it will be used as additional example in the next
sections. 

In practice, a log file is often a sequence of free-formed text lines
rather than a sequence of structured log entries. However, automatic
log parsing has been widely studied to decompose free-formed text
lines into structured log entries by accurately identifying fixed
parts (i.e., log message
templates)~\cite{7837916,8029742,messaoudi2018search,Zhu2019Tools,ELMASRI2020106276}. For
this reason, throughout the paper we assume that logs are given in a
structured form. Even though log parsing is not in the scope of this
work, we discuss how it affects the accuracy of \app in
Section~\ref{sec:discussion}.

\section{Operational Message Identification and Removal}\label{sec:technique}

The goal of \app is to identify operational templates in a set of
logs, and to remove from the logs the messages corresponding to these 
templates. For our running example shown in
Figure~\ref{fig:example-log}, this means that \app should identify the
templates \texttt{\textbf{ping}} and \texttt{\textbf{memory}} as
operational and remove the corresponding messages, highlighted in grey.

The intuition behind \app is that operational templates are
distinguishable from transactional templates by looking at two special
attributes, \emph{periodicity} and \emph{dependency}, of the messages
in the logs. For instance, in our running example, the 
\texttt{\textbf{ping}} template is distinguishable from the others because
its corresponding messages occur every 
second from the beginning until the end of the log (i.e., they have a
\emph{(global) periodicity} of one second). On the other hand, a
non-periodic, yet operational template like \texttt{\textbf{memory}}
is also distinguishable from the transactional templates
\texttt{\textbf{send}} and \texttt{\textbf{check}}, because
\begin{inparaenum}[(1)]
\item messages matching \texttt{\textbf{memory}} are randomly 
interleaved in the log (i.e., they do not have any \emph{periodicity});
\item one can see a \emph{dependency} between the occurrences of 
\texttt{\textbf{send}} and the occurrences of \texttt{\textbf{check}}
(i.e., in this case each occurrence of \texttt{\textbf{send}} is closely 
followed by an occurrence of \texttt{\textbf{check}}).
\end{inparaenum}

\begin{figure}
	\includegraphics[width=0.95\linewidth]{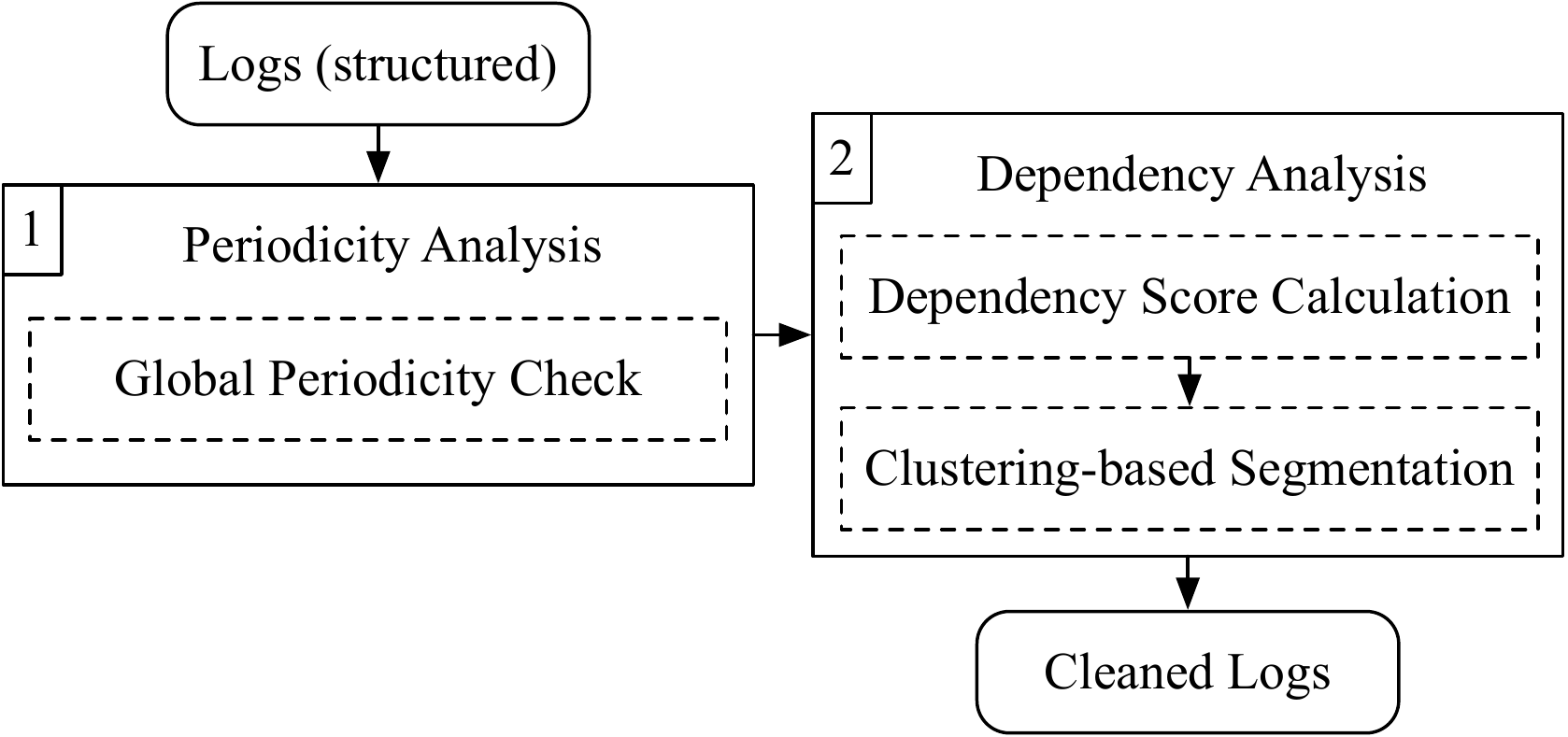}
	\caption{Overview of \app}
	\label{fig:overview}
	\Description{Overview of the steps included in \app}
\end{figure}

Based on these observations, the \app approach includes two main
steps, shown in figure~\ref{fig:overview}. The first step,
\emph{periodicity analysis}, aims at identifying ``globally periodic''
templates, i.e., heartbeat-like templates such as the
\texttt{\textbf{ping}} example above. The second step,
\emph{dependency analysis}, computes the degree of dependency among
templates, based on their co-occurrence, and relies
on a clustering-based heuristic to automatically partition operational
templates and transactional templates.
In the following subsections, we illustrate these two steps.

\subsection{Periodicity Analysis}\label{sec:periodicity-analysis}
The periodicity analysis filters out ``globally periodic'' templates,
like the \texttt{\textbf{ping}} template in our running example. Our
definition of ``global periodicity'' is as follows:
a template $t$ is \emph{globally periodic} for a set of logs $L$ if
$t$ occurs 
\begin{inparaenum}[(1)]
\item \emph{periodically} and
\item \emph{from the beginning to the end} of a log $l$, for all $l\in L$.
\end{inparaenum}
In our running example log $l_\mathit{org}$, the
\texttt{\textbf{ping}} template is globally periodic (for the given
log) since it periodically occurs every second from the
beginning to the end of the log. However, template
\texttt{\textbf{check}} is not globally periodic since it occurs at
\texttt{10:00:02}, \texttt{10:00:03}, and \texttt{10:00:07}, 
showing a non-periodic
behavior. Notice that if \texttt{\textbf{check}} had occurred at
\texttt{10:00:02}, \texttt{10:00:03}, and \texttt{10:00:04} (i.e.,
with a periodicity of one second), it would still not be considered
globally periodic because it would not have satisfied the second
condition for global periodicity.

\begin{algorithm}
\SetKwInOut{Input}{Input}
\SetKwInOut{Output}{Output}

\Input{Set of Logs $L$ \\ 
    Set of Templates $T$ \\
    Periodicity Deviation Threshold $\delta$, with $\delta \ge 0$}
\Output{Set of Logs $L_\mathit{cl}$}

Set of Templates $T_\mathit{gp} \gets \emptyset$\label{alg:periodic:init}\\
\ForEach{$t\in T$}{\label{alg:periodic:beginT}
    Boolean $\mathit{isGP} \gets$ \textbf{true}\label{alg:periodic:initgp}\\
    \ForEach{$l \in L$}{\label{alg:periodic:beginL}
        $\mathit{isGP} \gets \mathit{isGP}$ \textbf{and} 
		$\mathit{isPeriodicFromBeginToEnd}(t, l, \delta)$\label{alg:periodic:gptest}\\
    }\label{alg:periodic:endL}
    \If{$\mathit{isGP}$}{
        $T_\mathit{gp} \gets T_\mathit{gp} \cup \{t\}$\label{alg:periodic:addgp}\\
    }
}\label{alg:periodic:endT}
Set of Logs $L_\mathit{cl} \gets \mathit{removeMessagesOf}(T_\mathit{gp}, L)$\label{alg:periodic:rm}\\
\textbf{return} $L_\mathit{cl}$\\
\caption{Periodicity Analysis}
\label{alg:periodicity}
\end{algorithm}

Algorithm~\ref{alg:periodicity} shows the pseudo-code of the periodicity
analysis. It takes as input a set of logs $L$, a set of event
templates $T$ used in the messages in $L$, and a user-defined
periodicity threshold $\delta$, with $\delta \ge 0$; it returns a set
of cleaned logs $L_\mathit{cl}$, in which the globally periodic
messages (i.e., the messages with globally periodic templates) have
been removed.

The algorithm maintains a set of globally periodic templates $T_\mathit{gp}$, 
initially empty (line~\ref{alg:periodic:init}). $T_\mathit{gp}$ is populated by 
going through each template $t\in T$ 
(lines~\ref{alg:periodic:beginT}--\ref{alg:periodic:endT}), adding $t$ to 
$T_\mathit{gp}$ (line~\ref{alg:periodic:addgp}) if $t$ periodically occurs from 
the beginning to the end of a log $l$ with respect to $\delta$, for all $l\in L$ 
(lines~\ref{alg:periodic:beginL}--\ref{alg:periodic:endL}). At the end, 
$T_\mathit{gp}$ is used to remove the globally periodic messages from $L$, 
yielding $L_\mathit{cl}$ (line~\ref{alg:periodic:rm}). 

To check if a given template $t$ \emph{periodically occurs from the beginning to the 
end} of a given log $l$, we use the auxiliary function 
$\mathit{isPeriodicFromBeginToEnd}$, which analyzes the timestamps of the log entries 
of $t$ in $l$ (line~\ref{alg:periodic:gptest}).
This function computes the Mean Absolute Deviation (MAD) of the timestamp 
differences of the log entries of $t$ in $l$, and the Average of the
Timestamp Differences (ATD) of the log entries of $t$ in $l$. It then checks 
three conditions:
\begin{enumerate}
\item if the MAD value of the timestamp 
differences of the log entries of $t$ in $l$ is less than or equal to the
threshold $\delta$;
\item if the timestamp difference between the start of $l$ and the first entry of
  $t$ is at most ATD;
\item if the timestamp difference between the last entry of $t$ and the end of $l$
  is at least ATD.
\end{enumerate}	
The function returns \textit{true} only if $t$ satisfies the three
conditions in $l$.  Condition \#1 checks the periodicity of $t$ in
$l$; the threshold $\delta$ can be specified by an engineer depending
on the timestamp granularity in the logs, taking into account
timestamp skew caused by, for example, system overheads\footnote{In
  our experimentation, we notice that $\delta = 0.2$ is a reasonable
  threshold when the log timestamp granularity is in
  seconds.}. Conditions \#2 and \#3 check that $t$ occurs from the
beginning to the end of $l$, given that condition \#1 is satisfied.

For instance, in our running example log
$l_\mathit{org}$, the timestamp differences of the log entries of
the template \texttt{\textbf{ping}} are
$\langle 1, 1, \dots, 1 \rangle$, which leads to a MAD value of $0$;
the ATD of the log entries is one second. Also, the first log entry
of \texttt{\textbf{ping}} occurs at \texttt{10:00:01}, no later than
one second from the start of $l$; the last log entry of
\texttt{\textbf{ping}} occurs at \texttt{10:00:09}, not earlier than
one second from the end of $l$. Since template \texttt{\textbf{ping}}
satisfies the three conditions in the example log, the auxiliary
function returns \textbf{true}. On the other hand, the function
returns \textit{false} for templates \texttt{\textbf{check}} and
\texttt{\textbf{memory}} because they do not satisfy the first
condition. It also returns \textit{false} for template
\texttt{\textbf{send}} because it only has two occurrences in the log
and the periodicity cannot be checked\footnote{We require at least
  three log entries (i.e., at least two timestamp differences) to check the
  periodicity.}.
As a result, the periodicity analysis removes the log entries with
template \texttt{\textbf{ping}} in $l_\mathit{org}$; the resulting log is
shown in Figure~\ref{fig:example-log} as $l_\mathit{inter}$.

\subsection{Dependency Analysis}\label{sec:dep-analysis}
The dependency analysis removes non-periodic operational templates,
like the \texttt{\textbf{memory}} template in our running example,
which are randomly interleaved with transactional templates. To do
this, the dependency analysis computes the degree of dependency
(hereafter called \emph{dependency score}) among
templates based on the co-occurrences of the templates in logs, and
then uses the computed dependency scores to distinguish operational
templates and transactional templates.

The underpinning idea of the dependency analysis comes from the
characteristics of transactional and operational events in a
system. Since the occurrence of transactional events reflects the flow
of the functional behavior of a system, a transactional event has a
dependency either with its predecessor or successor event in the flow.
On the contrary, an operational event can occur independently from the
flow of a system, since its occurrence reflects the system state
(e.g., the memory usage) instead of its functional behavior. 
Assuming there is a way
to measure, in the logs recorded during the execution of the system,
the degree of dependency between templates, \emph{we expect
operational templates to have a much lower dependency score on other
templates than transactional templates}. Therefore, we use a
clustering-based heuristic to automatically partition operational
templates and transactional templates based on the dependency score of
each template on the others.

\begin{algorithm}
\SetKwInOut{Input}{Input}
\SetKwInOut{Output}{Output}

\Input{Set of Logs $L$ \\
    Set of Templates $T$}
\Output{Set of Logs $L_\mathit{cl}$}

Map from $T$ to Set of Reals $\mathit{mScore}$ \label{alg:dep:init}\\
\ForEach{$x\in T$}{\label{alg:dep:beginX}
    $\mathit{mScore}[x] \gets 0$ \\
    \ForEach{$y\in T \setminus \{x\}$}{\label{alg:dep:beginY}
        $\mathit{mScore}[x] \gets \max(\mathit{mScore}[x], \mathit{dScoreCalc}(x, y, L))$\label{alg:dep:dScore}\\
    }\label{alg:dep:endY}
}\label{alg:dep:endX}
Set of Templates $T_\mathit{op} \gets \mathit{clusterBasedSegment}(T, mScore)$\label{alg:dep:seg}\\
Set of Logs $L_\mathit{cl} \gets \mathit{removeMessagesOf}(T_\mathit{op}, L)$\label{alg:dep:rm}\\
\textbf{return} $L_\mathit{cl}$\\

\caption{Dependency Analysis}
\label{alg:dependency}
\end{algorithm}

Algorithm~\ref{alg:dependency} shows the pseudo-code of the dependency analysis. 
It takes as input a set of logs $L$ and a set of templates $T$; it 
returns a set of cleaned logs $L_\mathit{cl}$, in which the operational messages 
(i.e., the messages having operational templates) have been removed.

For each template $x\in T$, the algorithm determines the maximum value
of the dependency score (the value $\mathit{mScore}[x]$ for the key
$x$ in the associative array $\mathit{mScore}$), by computing the
individual dependency scores of $x$ on the other templates
$y \in T \setminus \{x\}$ in $L$
(lines~\ref{alg:dep:beginX}--\ref{alg:dep:endX}); this last step is
done by the \textit{dScoreCalc} function, described in detail in
\S~\ref{sec:dScore}. Using the calculated $\mathit{mScore}$, the
algorithm calls the \textit{clusterBasedSegment} function (described
in detail in \S~\ref{sec:clustering}) to identify the set of
operational templates $T_\mathit{op}$ from $T$
(line~\ref{alg:dep:seg}). The algorithm ends by returning the set of
cleaned logs $L_\mathit{cl}$, obtained (line~\ref{alg:dep:rm}) by
removing the operational messages from $L$ based on the operational
templates in $T_\mathit{op}$.

\subsubsection{Log-based Dependency Score}\label{sec:dScore}
To measure the dependency score of a template $x$ on another template
$y$ for a set of logs $L$, we consider not only the dependency for
which $x$ could be a \emph{cause} of $y$ in $L$ but also the
dependency for which $x$ could be a \emph{consequence} of $y$ in $L$.
More precisely, we define the \emph{forward dependency score} of $x$
on $y$ for $L$, denoted with $\mathit{dScore}_f(x, y, L)$, as a
measure of how likely an occurrence of $x$ is \emph{followed} by an
occurrence of $y$ (i.e., $x$ is a cause of $y$) throughout $L$.
Similarly, the \emph{backward dependency score} of $x$ on $y$ for $L$,
denoted with $\mathit{dScore}_b(x, y, L)$, is a measure of how likely
an occurrence of $x$ is \emph{preceded} by an occurrence of $y$ (i.e.,
$x$ is a \emph{consequence} of $y$) throughout $L$. Since
$\mathit{dScore}_b(x, y, L)$ is equivalent to
$\mathit{dScore}_f(x, y, \mathit{rev}(L))$, where $\mathit{rev}(L)$ is
the set containing the reversed logs of $L$, below we only present the
algorithm to compute the forward dependency score.

First, we introduce the concept of a log entry occurring 
after another one. More formally, given a log entry $e_x$ of a
template $x$ in a log $l$, we say that a log entry $e_y$ of a template
$y$ is the \emph{first-following} log entry for $e_x$ in $l$ if $e_y$
is the first log entry of $y$ between $e_x$ and the next log entry of
$x$ in $l$. For instance, in our
running example log $l_\mathit{inter}$, the log entries of the
\texttt{\textbf{memory}} template are $e_{2'}$, $e_{5'}$, $e_{6'}$,
and $e_{9'}$ while the log entries of the \texttt{\textbf{check}}
template are $e_{3'}$, $e_{4'}$, and $e_{8'}$. The log entry $e_{3'}$
(of template \texttt{\textbf{check}}) is the first-following log entry
for $e_{2'}$ (of template \texttt{\textbf{memory}}),
because $e_{3'}$ is the first log entry of \texttt{\textbf{check}}
between $e_{2'}$ and $e_{5'}$. Similarly, $e_{8'}$ is the
first-following log entry for $e_{6'}$. However, there is no
first-following log entry for $e_{5'}$ because there is no log entry of
\texttt{\textbf{check}} between $e_{5'}$ and $e_{6'}$. Also, $e_{9'}$
has no first-following log entry because it occurs at the end of the
log.

However, simply checking whether there is a first-following log entry
is not enough to compute the dependency score, because it does not
consider \emph{how close} the two log entries are. To take into
account the distance between log entries, we define the
\emph{co-occurrence score} between two log entries $e_x$ and $e_y$ in
a log $l$, denoted with $\mathit{cScore}(e_x, e_y, l)$, as
$ \mathit{cScore}(e_x, e_y, l) = \frac{1}{\mathit{distance}(e_x, e_y,
 l)}$, where $\mathit{distance}(e_x, e_y, l)$ is the difference of
the indexes between $e_x$ and $e_y$ in $l$. For our running example
with $l = l_\mathit{inter}$, we have
$\mathit{cScore}(e_{2'}, e_{3'}, l_\mathit{inter}) = 1$ because
$\mathit{distance}(e_{2'}, e_{3'}, l_\mathit{inter})=1$, 
and
$\mathit{cScore}(e_{6'}, e_{8'}, l_\mathit{inter}) = 0.5$ because 
$\mathit{distance}(e_{6'}, e_{8'}, l_\mathit{inter})=2$.
When an entry $e_x$ has no first-following log entry, 
as it is the case for $e_{5'}$ and $e_{9'}$ in the above example,
we have $\mathit{cScore}(e_x, \mathit{NaE}, l) = 0$ where $\mathit{NaE}$ indicates ``Not an Entry''.
We can then compute the dependency score between two templates $x$ and $y$
through a set of logs $L$ as the average of the 
$\mathit{cScore}$ values of all log entries of $x$ with its first-following 
log entries of $y$. 
More formally, we have:
\begin{displaymath}
\mathit{dScore}_f(x, y, L) = 
\frac{\sum_{l \in L}{\sum_{e_x \in E_{x,l}}{\mathit{cScore}(e_x, e_y, l)}}}{n}
\end{displaymath}
where $E_{x,l}$ is the set of log entries of $x$ in $l$,
$e_y$ is the first-following entry of $y$ for $e_x$ in $l$,
and $n$ is the total number of log entries of $x$ in $L$.
In this way, we measure how likely and how closely an occurrence of $x$ is 
followed by an occurrence of $y$ throughout $L$.
A value of $\mathit{dScore}_f(x, y, L) = 1$ indicates that $x$ always
immediately causes $y$ in the logs in $L$, 
while a value $\mathit{dScore}_f(x, y, L) = 0$ indicates that $x$ cannot cause $y$ in the logs in $L$.
For the running example above, we have:
\begin{align*}
\mathit{dScore}_f&(\texttt{\textbf{memory}}, \texttt{\textbf{check}},\{l_\mathit{inter}\}) \\
&= \frac{\sum_{l \in \{l_\mathit{inter}\}}{\sum_{e_x \in E_{x,l}}{\mathit{cScore}(e_x, e_y, l)}}}{4} \\
&= \frac{ \sum_{e_x \in \{e_{2'}, e_{5'}, e_{6'}, e_{9'}\}} {\mathit{cScore}(e_x, e_y, l_\mathit{inter})} }{4} \\
&= \frac{1}{4} \times \left[ \mathit{cScore}(e_{2'}, e_{3'}, l_\mathit{inter}) + \mathit{cScore}(e_{5'}, \mathit{NaE}, l_\mathit{inter}) \right. \\
&\qquad \left. + \text{ } \mathit{cScore}(e_{6'}, e_{8'}, l_\mathit{inter}) + \mathit{cScore}(e_{9'}, \mathit{NaE}, l_\mathit{inter}) \right] \\
&= \frac{1}{4} \times \left[ 1 + 0 + 0.5 + 0 \right] = 0.375
\end{align*}

We recall that the values of the dependency score are used in
Algorithm~\ref{alg:dependency} to compute the maximum dependency score
$\mathit{mScore}$ of each template; for our running example log
$l_\mathit{inter}$ with the set of templates
$T = \{ \texttt{\textbf{send}}, \texttt{\textbf{check}},
\texttt{\textbf{memory}} \}$, we have
$\mathit{mScore}[\texttt{\textbf{send}}] = 0.75$,
$\mathit{mScore}[\texttt{\textbf{check}}] = 0.67$, and
$\mathit{mScore}[\texttt{\textbf{memory}}] = 0.5$.

\subsubsection{Clustering-based Segmentation}\label{sec:clustering}
As mentioned earlier, the (maximum) dependency score
$\mathit{mScore}$ value of operational templates is likely to be less than
that of transactional templates. Furthermore, in our preliminary
experiments, we observed that the gap among the $\mathit{mScore}$ 
values for operational templates is often smaller than the gap
between the highest $\mathit{mScore}$ value of an operation
template and the lowest $\mathit{mScore}$ value of a transactional
template. This suggests that the set of operational templates could
form a cluster based on $\mathit{mScore}$. Therefore, we propose a
heuristic to partition operational templates and transactional
templates using clustering.

We first generate multiple clusters of templates based on the value
$\mathit{mScore}$. The number of clusters can be more than two because
transactional templates often lead to multiple clusters. Since the
number of clusters is not known in advance, we use the Mean-Shift
clustering algorithm~\cite{comaniciu02:_mean}.
Among the generated clusters, the cluster with the smallest 
$\mathit{mScore}$ value is assumed to be the one of operational templates. 

For instance, if we apply the clustering-based segmentation heuristic
to our running example, with
$T = \{ \texttt{\textbf{send}}, \texttt{\textbf{check}},
\texttt{\textbf{memory}} \}$ and using the $\mathit{mScore}$ values
computed in \S~\ref{sec:dScore}, the clustering algorithm will
generate three clusters $c_1 = \{ \texttt{\textbf{send}} \}$,
$c_2 =\{ \texttt{\textbf{check}} \}$, and
$c_3 = \{ \texttt{\textbf{memory}} \}$. Since $c_3$ has the smallest
$\mathit{mScore} = 0.5$, the \texttt{\textbf{memory}} template in
$c_3$ is identified as operational.

After identifying operational templates with the clustering-based
segmentation heuristic, the dependency analysis algorithm ends with
the removal of the log entries containing one of the identified 
operational templates. In the case of our running example log 
$I_\mathit{inter}$, this means removing the entries with template
\texttt{\textbf{memory}}; the final, cleaned version of the log is the one
without any of the operational messages highlighted in grey 
in Figure~\ref{fig:example-log}.

\section{Evaluation}\label{sec:eval}
In this section, we report on the experimental evaluation of the
effectiveness of \app in removing operational event
templates and its effect on model inference.

More precisely, we assess the accuracy of \app in terms of removing
operational messages, to determine its suitability as a
pre-processing step before model inference.
Furthermore, we want to investigate the impact of \app on model inference
in terms of cost (i.e., execution time), when used to pre-process
the input logs before inferring models. Since \app reduces the size of the
logs given as input to a model inference tool, we expect a cost
reduction of the model inference process as well.
We also want to analyze the impact of \app on model inference in terms
of accuracy of the inferred model: since \app removes
``noise'' from logs (in the form of operational log messages), we
expect an improvement of the accuracy of the inferred models.

Summing up, we investigate the following research questions:
\begin{itemize}
\item RQ1: \emph{What is the accuracy of \app in removing
  operational event templates?} (subsection~\ref{sec:RQ1})
\item RQ2: \emph{What is the impact of \app on model inference in
    terms of cost (execution time)?} (subsection~\ref{sec:RQ2})
\item RQ3: \emph{What is the impact of \app on model inference in
    terms of accuracy of the inferred models?}  (subsection~\ref{sec:RQ3})
\end{itemize}

\subsection{Benchmark and Settings}\label{sec:benchmark}
To evaluate \app, we assembled a benchmark composed of proprietary and
non-proprietary logs obtained from the execution of different types
of systems.  Table~\ref{table:subjects} lists the systems we included
in the benchmark and statistics about the corresponding logs: the number of logs
(column \emph{\# Logs}), the number of templates (column \emph{\#
  Templates}), and the total number of log entries (column \emph{\# Entries}).

\begin{table}
\footnotesize
 \caption{Subject Systems and Logs}\label{table:subjects}
 \begin{tabular}{llrrr}
 \toprule
Type & System & \# Logs & \# Templates & \# Entries \\
 \midrule
\multirow{2}{*}{Proprietary} & \texttt{SYS1} & 120 & 17 & 22162 \\
& \texttt{SYS2} & 36 & 5 & 569 \\
\midrule
\multirow{11}{*}{Non-Proprietary} & \texttt{CVS} & 3963 & 15 & 31977 \\
& \texttt{Lucane} & 1000 & 16 & 14693 \\
& \texttt{RapidMiner} & 1340 & 18 & 25440 \\
& \texttt{SSH} & 3624 & 9 & 68436 \\
& \texttt{DatagramSocket} & 1000 & 28 & 16061 \\
& \texttt{MultiCastSocket} & 1000 & 15 & 12801 \\
& \texttt{Socket} & 2304 & 41 & 49724 \\
& \texttt{Formatter} & 1000 & 7 & 7476 \\
& \texttt{StringTokenizer} & 1000 & 6 & 5139 \\
& \texttt{TCPIP} & 2350 & 10 & 40966 \\
& \texttt{URL} & 1000 & 16 & 21845 \\
\bottomrule
\end{tabular}
\end{table}

The proprietary logs have been recorded during the execution of
two subsystems---named \texttt{SYS1} and \texttt{SYS2} for
confidentiality reasons---of the ground control system operated by our
industrial partner in the satellite domain. We selected these two
subsystems because the engineers of our partner indicated them as the top
producers of ``noisy'' logs, which are difficult to analyze and process (and thus
could benefit most from the application of \app). However, these
proprietary logs are not documented, and the corresponding systems are
composed of black-box, 3rd-party components. Because of these reasons,
even the engineers of our industrial partner cannot fully interpret
the meaning of the log event templates and
classify them as transactional or operational templates. Furthermore,
since these logs are obtained from the execution of a real system, we
cannot  conduct controlled experiments using them.

To overcome these limitations and to support open science, we included
in our benchmark logs generated from 11 publicly available system
models (in the form of Finite State Machines - FSM), previously
proposed in the
literature~\cite{postel1981transmission,poll2007verifying,4023976,5609576,LO20122063}.
These FSM models, by design, represent the functional
behavior of a system; the logs generated from them are purely
transactional.
We used the methodology proposed by \citet{nimrod} to generate logs
from these models, using the publicly available trace generator by
\citet{LO20122063}, configured to provide state coverage of four
visits per state and a minimum of 1000 logs.  Since the log entries of
the generated logs have no actual timestamps, we used the indexes of
the log entries as their timestamps.

We conducted our evaluation on a high-performance computing platform, using one
of its quad-core nodes running CentOS 7 on a \SI{2.4}{\giga\hertz} Intel Xeon E5-2680 v4
processor with \SI{4}{\giga\byte} memory.

\subsection{Accuracy of \app}\label{sec:RQ1}
To answer RQ1, we assess how accurately \app removes operational
templates while preserving transactional templates.

\subsubsection{Methodology}\label{sec:RQ1:setup}
We measured the accuracy of \app in terms of \emph{recall} and
\emph{specificity}, where recall indicates how accurately operational
templates are removed and specificity indicates how accurately
transactional templates are preserved.
More specifically, we say that if an operational template is correctly
removed by \app, it is classified as True Positive (TP); otherwise, it is
classified as False Negative (FN). Similarly, if a transactional template is
correctly preserved by \app, it is classified as True Negative (TN); otherwise,
it is classified as False Positive (FP). Based on the classification results for
all templates, we have
$\mathit{Recall}=\tfrac{|\mathit{TP}|}{|\mathit{TP}|+|\mathit{FN}|}$
and $\mathit{Specificity}=\tfrac{|\mathit{TN}|}{|\mathit{TN}|+|\mathit{FP}|}$.
Both recall and specificity values range from 0 to 1, where 1
indicates best and 0 indicates worst.

Since the logs of the proprietary systems are not documented (see
discussion in subsection~\ref{sec:benchmark}) and no ground truth is
available for them, we excluded them in our experiments for RQ1.  As
for the logs generated from the publicly available system models,
since they are purely transactional, we randomly injected operational
messages into the logs. More specifically, we randomly generated the
log entries of $n=5$ operational templates and then randomly injected
them into the logs by randomizing the timestamps of the generated
entries. In this way, logs contain transactional messages randomly
interleaved with operational messages.
To better understand the accuracy of \app with respect to the
proportion of operational messages, we varied the Noise Rate (NR), i.e.,
the number of injected operational log entries over the total number of log
entries, from 0.1 to 0.9 in steps of 0.1.

Note that though \app is not randomized, the noise injection procedure includes
randomness. Thus, we ran \app 30 times and measured the average recall and
specificity.

We did not compare \app to existing work, such as the filtering
technique presented by \citet{8030620}, because neither the
implementation nor a clear algorithm of the filtering technique are
available in the article.  Instead, we provide a detailed technical
comparison in Section~\ref{sec:related-work}.

\subsubsection{Results}\label{sec:RQ1:results}

\begin{figure*}
	\centering
	\input{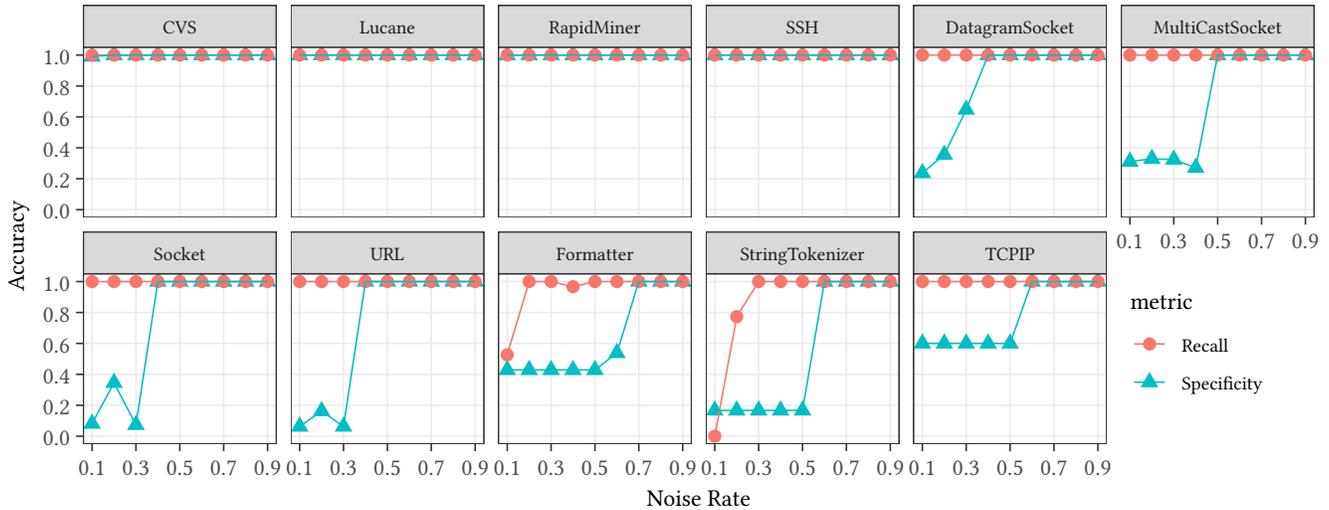}\caption{Relationship between Noise Rate and Accuracy of \app}
	\label{fig:NR-A}
	\Description{Relationship between Noise Rate and Accuracy of \app}
\end{figure*}

Figure~\ref{fig:NR-A} shows the recall and specificity of \app as a function of
the NR for each subject system.

On average across all cases, recall is 0.98 and specificity is 0.81.
We can see that the results for the four systems \texttt{CVS},
\texttt{Lucane}, \texttt{RapidMiner}, and \texttt{SSH} (hereafter
called \texttt{Type-A}) are distinct from the results for the other
seven systems (hereafter called \texttt{Type-B}).  For the four
\texttt{Type-A} systems, both recall and specificity are very high
(i.e., always 1 except 0.99 for \texttt{CVS} when NR = 0.1) regardless
of the NR. This means that \app, for this type of systems, is nearly
perfect at removing operational templates and at preserving
transactional templates, regardless of the proportion of operational
messages in the logs.  On the other hand, for the seven \texttt{Type-B}
systems, recall is generally high (0.97 on average) while specificity
is not (0.69 on average). This means that \app is good at removing
operational templates in general, but often \app incorrectly removes
some transactional templates.

To investigate the reason why the accuracy is different between
\texttt{Type-A} and \texttt{Type-B} systems, we took a closer look at
the underlying models from which the logs were generated. More
specifically, for each FSM model (with input alphabet $\Sigma$), we first measured
the \emph{event diversity score (\textit{eDiv-Score})} of each input symbol
$\sigma \in \Sigma$ (i.e., a transactional event), defined as the ratio between
\begin{inparaenum}[(1)]
  \item the number of unique input symbols on the outgoing transitions
  of all states that can be reached upon the occurrence of $\sigma$, and
  \item the cardinality of the input alphabet of the FSM, i.e., $|\Sigma|$.
\end{inparaenum}
The \textit{eDiv-Score} of $\sigma$ ranges between 0 to 1, indicating the proportion of
transactional events that can randomly occur immediately after the occurrence of
$\sigma$. We then measured the \emph{system diversity score
(\textit{sDiv-Score})} of a system model, defined by the average \textit{eDiv-Score}
for all transactional events of the system model. Note that
the \textit{sDiv-Score} is a characteristic of the system, not of the
corresponding logs; however, it is reflected in the sequence of events
recorded in a log. The resulting \textit{sDiv-Score} for the
\texttt{Type-A} systems ranges between 0.10 (\texttt{Lucane}) and 0.41
(\texttt{SSH}), while the \textit{sDiv-Score} for the \texttt{Type-B}
systems ranges between 0.56 (\texttt{TCPIP}) and 0.93
(\texttt{URL}). This means that, in the logs of the \texttt{Type-B}
systems, on average, more than half of all transactional templates can randomly
occur immediately after any transactional template. This
random interleaving of transactional messages affects
dependency analysis, and is the reason for which
\app is not very accurate for \texttt{Type-B} systems.

In practice, we expect the \textit{sDiv-Score} to be inversely
proportional to the quality of the logging statements, where quality
can be defined in terms of  ``what to log''~\cite{He2018ASE} and
``where to log''~\cite{Zhu:2015:LLH:2818754.2818807}.
For example, a developer might decide to record a file opening event
with fine-grained event templates that record the distinct
read/write modes, such as
``\texttt{\textbf{open}} \texttt{file} \texttt{\textbf{(read)}}'',
``\texttt{\textbf{open}} \texttt{file} \texttt{\textbf{(write)}}'',
and ``\texttt{\textbf{open}} \texttt{file} \texttt{\textbf{(append)}}''.
This is good if the functional behavior of the system being logged
varies depending on the distinct read/write modes.
However, if this is not the case, a coarse-grained log message
template, such as ``\texttt{\textbf{open}} \texttt{file}'', is better.
In any cases, if the granularity of the logging statements reflects
the functional behavior of the system, the \textit{sDiv-Score}
will be reasonably low.

Nevertheless, \app achieves a perfect specificity for
\texttt{Type-B} systems when NR $\ge$ 0.7. This means that, if more
than 70\% of all log messages are operational, \app is able to
perfectly preserve transactional templates even when transactional
messages are randomly interleaved. At first, this might seems
counter-intuitive because it is likely that one would no longer be
able to distinguish between operational and transactional messages if
messages of both types are randomly interleaved. However, the more
operational messages, the more occurrences (of each operational
template) that are randomly interleaved.  This leads to a decrease of the
dependency scores of the operational templates, an increase of the
dependency score gap between the operational templates and the
transactional templates, and ultimately an increase of the specificity
of \app.

To summarize, the answer to RQ1 is that \app correctly removes
operational messages and correctly preserves transactional messages
with pinpoint accuracy, regardless of the noise level in the logs, for
systems with high-quality logs. On the other hand, for systems with
low-quality logs, \app may incorrectly remove some transactional
messages depending on the noise level in the logs, achieving an
average accuracy of 69\%. Nevertheless, for such systems, the accuracy
is always 100\% when the proportion of the noise level exceeds 70\% of all
log messages.
Furthermore, the impact of such information loss on model inference remains
to be investigated. Incorrectly removing some of the transactional messages may not
necessarily lead to (significant) negative effects.

Building on the above results, in the next research questions, we will
investigate the impact of log cleaning with \app on model inference,
including the cases when \app incorrectly removes transactional
messages due to low-quality logs.

\subsection{Impact of \app on Model Inference Cost (Execution Time)}\label{sec:RQ2}
To answer RQ2, we compare the execution time of model
inference on
\begin{inparaenum}[(1)]
	\item unmodified, original logs containing operational messages and
	\item logs that have been pre-processed using \app,
\end{inparaenum}
while accounting for the execution time of pre-processing as well.

\subsubsection{Methodology}
We selected MINT~\cite{minttool} as model inference tool because it is
a state-of-the-art, publicly
available tool, known to be accurate.

We prepared the logs to be used for model inference as follows.
We injected operational messages into the logs of the 11 publicly
available systems using the same procedure indicated above for
RQ1\footnote{We recall that operational message injection does not increase the number of
logs.}. Since the results of RQ1 show that the accuracy of \app varies
as a function of NR, we selected two representative settings
(NR=0.3 and NR=0.7) to see how the cost of model
inference varies with 0 .the accuracy of \app. We did not
modify the logs of the two proprietary systems since these logs
already contained both operational and transactional messages.

Given that the problem of inferring a minimal FSM is
NP-complete~\cite{biermann1972synthesis}, and that the limitations of MINT in terms of
scalability are well-known~\cite{wang2016scalable,LUO201713}, we conducted some
preliminary experiments for identifying the number of logs that MINT could
process with a 1-hour timeout for each of the subject systems. Through these
preliminary experiments, we found that MINT could process in the given timeout:
\begin{inparaenum}[(a)]
  \item up to 100 logs  for each non-proprietary system,
  \item up to 20 logs for  the proprietary system \texttt{SYS1}, and
  \item all 36 logs for  the proprietary system \texttt{SYS2}.
\end{inparaenum}
Based on these preliminary results, for each non-proprietary system in
our benchmark, we randomly selected 100 logs to be used for model inference;
similarly, for \texttt{SYS1}, we randomly selected 20 logs;
for \texttt{SYS2}, we used all 36 logs available.

For each system in our benchmark, let $L_\mathit{org}$ be the set of
original logs selected for model inference,
containing both operational and transactional messages of the system. We ran
\app on $L_\mathit{org}$ to generate a set of cleaned logs $L_\mathit{cl}$ and
measured the execution time of \app. We then ran MINT on both
$L_\mathit{org}$ and $L_\mathit{cl}$, measuring the execution time. We
calculated \emph{Speed-Up} as
$\mathit{SU}_\mathit{cl}=\frac{M_\mathit{org}}{M_\mathit{cl} + \mathit{LgCl}}$,
where $M_x$ indicates the execution time of MINT on $L_x$ and $\mathit{LgCl}$
indicates the execution time of \app (on $L_\mathit{org}$ to generate $L_\mathit{cl}$).
$\mathit{SU}_\mathit{cl}$ indicates how much faster
model inference is when the input logs are pre-processed using \app.

However, since \app reduces the size (i.e., the total number of log entries) of
$L_\mathit{cl}$ while removing operational messages, comparing the execution
time of MINT for processing $L_\mathit{cl}$ and $L_\mathit{org}$ would not
clearly show whether any difference is mainly due to the
removal of operational messages or to the reduction in size of the logs.
To avoid such an issue, we built a new baseline set of logs
$L_\mathit{cl+}$, whose size is the same as that of
$L_\mathit{org}$, while the noise level is the same as that
of $L_\mathit{cl}$. We generated $L_\mathit{cl+}$ starting from $L_\mathit{cl}$, so
that both incorrectly preserved operational templates (i.e., noise) and incorrectly
removed transactional templates (i.e., information-loss) in $L_\mathit{cl}$ are
the same in $L_\mathit{cl+}$. Furthermore, we preserved the correct sequences of
transactional messages, and only changed the position of operational
messages. As
a result, $L_\mathit{cl+}$ represents a set of cleaned logs generated from
$L_\mathit{org}$ using \app, with the same size as $L_\mathit{org}$.
We computed $\mathit{SU}_\mathit{cl+}$ in the same way as
$\mathit{SU}_\mathit{cl}$, replacing $L_\mathit{cl}$ with
$L_\mathit{cl+}$. We generated $L_\mathit{cl+}$ (and computed
$\mathit{SU}_\mathit{cl+}$) only for the non-proprietary systems, since
the generation step requires to know the ground truth about the
operational messages.

To account for the   randomness in the generation of $L_\mathit{org}$ and the
injection of operational messages, we repeated the experiment 30 times and
computed the average results. Furthermore, we applied the non-parametric
Wilcoxon signed-ranks test to assess the statistical significance of the
difference in the execution time of MINT between $L_\mathit{org}$ and
$L_\mathit{cl}$ and between $L_\mathit{org}$ and $L_\mathit{cl+}$,
accounting for the execution time of \app.

\subsubsection{Results}
Table~\ref{table:RQ2} shows the results of the impact of \app on model
inference cost, grouped by NR and system. Column $M_x$ indicates the
execution time of MINT on $L_x$ where
$x \in \{\mathit{org}, \mathit{cl}, \mathit{cl+}\}$; column
$\mathit{LgCl}$ indicates the execution time of \app; columns
$\mathit{SU}_\mathit{cl}$ and $\mathit{SU}_\mathit{cl+}$ indicate the
speed-up values.  All the values are the average results achieved
across 30 executions of \app. In all cases, the execution time
differences between $L_\mathit{org}$ and $L_\mathit{cl}$ (or
$L_\mathit{cl+}$) are statistically significant ($p\text{-value} < 0.01$).

\begin{table}
\footnotesize
\caption{Impact of \app on Model Inference Cost}\label{table:RQ2}
\begin{tabular}{llrrrrrr}
\toprule
	&	& \multicolumn{4}{c}{Execution time (s)}  & \multicolumn{2}{c}{Speed-up} \\
\cmidrule(r){3-6}\cmidrule(r){7-8}
NR                    & System          & $M_\mathit{org}$  & $M_\mathit{cl}$  & $M_\mathit{cl+}$ & $\mathit{LgCl}$ & $\mathit{SU}_\mathit{cl}$ & $\mathit{SU}_\mathit{cl+}$ \\
\midrule
\multirow{2}{*}{-}      & \texttt{SYS1}            & 2008   & 13.9  & NA    & 0.3 & 161.7 & NA    \\
                        & \texttt{SYS2}            & 0.9    & 0.7   & NA    & 0   & 1.3   & NA    \\
\midrule
\multirow{11}{*}{0.3}   & \texttt{CVS}             & 89.9   & 1.7   & 1.7   & 0.3 & 47.4  & 48.4  \\
                        & \texttt{Lucane}          & 91.9   & 2.9   & 2.9   & 0.6 & 27.8  & 27.5  \\
                        & \texttt{RapidMiner}      & 499.8  & 4.8   & 5.4   & 0.8 & 95.7  & 86.7  \\
                        & \texttt{SSH}             & 414.5  & 97.7  & 119.7 & 0.4 & 6.6   & 5.4   \\
                        & \texttt{DatagramSocket}  & 444.1  & 12.2  & 14    & 1.3 & 135.9 & 127.7 \\
                        & \texttt{MultiCastSocket} & 268.1  & 0.5   & 0.6   & 0.5 & 284.7 & 256.1 \\
                        & \texttt{Socket}          & 642.4  & 13.4  & 15.8  & 2.9 & 177.4 & 167.6 \\
                        & \texttt{Formatter}       & 54     & 7.5   & 12.3  & 0.2 & 69.6  & 61.6  \\
                        & \texttt{StringTokenizer} & 44.4   & 0.6   & 1.1   & 0.1 & 86.5  & 82.4  \\
                        & \texttt{TCPIP}           & 360.3  & 16.1  & 26.4  & 0.4 & 63.0  & 45.7  \\
                        & \texttt{URL}             & 1073.1 & 0.3   & 0.4   & 0.8 & 946.7 & 882.1 \\
\midrule
\multirow{11}{*}{0.7}   & \texttt{CVS}             & 124.4  & 1.6   & 1.8   & 0.6 & 56.2  & 51.9  \\
                        & \texttt{Lucane}          & 257.3  & 2.8   & 3.1   & 1.1 & 68.9  & 64.1  \\
                        & \texttt{RapidMiner}      & 440.4  & 5.5   & 6.5   & 1.5 & 67.5  & 57.6  \\
                        & \texttt{SSH}             & 508.9  & 103.3 & 138   & 0.7 & 7.9   & 5.2   \\
                        & \texttt{DatagramSocket}  & 274.9  & 192.8 & 223   & 2.3 & 1.5   & 1.3   \\
                        & \texttt{MultiCastSocket} & 176.1  & 68.3  & 74.2  & 0.9 & 2.8   & 2.5   \\
                        & \texttt{Socket}          & 610.3  & 326.1 & 367.6 & 5.4 & 1.8   & 1.7   \\
                        & \texttt{Formatter}       & 103.3  & 7.9   & 11.2  & 0.3 & 13.9  & 11.7  \\
                        & \texttt{StringTokenizer} & 56     & 6.4   & 6.7   & 0.2 & 9.5   & 9.0   \\
                        & \texttt{TCPIP}           & 383.6  & 46.7  & 62    & 0.7 & 10.5  & 7.7   \\
                        & \texttt{URL}             & 735.5  & 477.7 & 472.1 & 1.5 & 1.8   & 2.1   \\
\bottomrule
\end{tabular}
\end{table}

We can see that $\mathit{SU}_\mathit{cl}$ is similar to
$\mathit{SU}_\mathit{cl+}$ in all cases, implying that the cost
reduction of model inference is mainly due to the removal of
operational messages using \app, not to the size reduction of input
logs. However, $\mathit{SU}$ varies greatly from system to system,
even within the same NR group; for example, when NR = 0.7,
$\mathit{SU}_\mathit{cl}$ ranges between 1.8 and 68.9 (with a mean of
22.0 and a standard deviation of 27.6). This is due to the intrinsic
characteristics of the system being logged (reflected in the size of
the model to infer), which affect the model inference process.

We remark that the simpler the model to infer, the shorter the model inference time.
Thus, when \app incorrectly removes transactional messages in addition to operational messages
(e.g., for \texttt{Type-B} systems when NR = 0.3), the remaining
transactional messages in $L_\mathit{cl}$ lead to the inference of a less complex
model, resulting in a very high $\mathit{SU}$; for example, for the same \texttt{URL} system,
$\mathit{SU}_\mathit{cl} = 946.7$ when NR = 0.3 (when \app incorrectly removes
additional transactional messages) whereas $\mathit{SU}_\mathit{cl} = 1.8$ when
NR = 0.7 (when \app perfectly preserves transactional messages).

To summarize, the answer to RQ2 is that pre-processing the input logs
using \app reduces the cost (in terms of execution time) of model
inference, with a speed-up ranging from 1.5 to 946.7 depending on the
characteristics of the system.  Even when removing the effect of the log
size reduction, the speed-up with \app pre-processing ranges from 1.3
to 882.1.  Though there is significant variation across systems and NR
values, \app is beneficial in all cases.

Finally, we want to remark that, for large logs that are commonly
encountered in industrial contexts, the magnitude of cost reduction
in model inference will be more significant than in our experiments.
For example, MINT takes more than
10 hours when we use all 120 logs for the proprietary system \texttt{SYS1},
whereas it takes less than one minute when the same 120 logs are pre-processed using \app.

\subsection{Impact of \app on Model Inference Accuracy}\label{sec:RQ3}
To answer RQ3, we compare the accuracy of the models inferred from
\begin{inparaenum}[(1)]
  \item unmodified, original logs containing operational messages, and
  \item logs that have been pre-processed using \app.
\end{inparaenum}

\subsubsection{Methodology}\label{sec:RQ3:methodology}
As done for RQ2, we used MINT as model inference tool. Also, we
prepared $L_\mathit{org}$ for each system in our benchmark in the same
way we did for RQ2.  We ran \app to generate $L_\mathit{cl}$ from
$L_\mathit{org}$, and then ran MINT on both $L_\mathit{org}$ and
$L_\mathit{cl}$ to infer a model and then measure model inference accuracy.

Following previous model inference studies~\cite{walkinshaw2016inferring,
mariani2017gk, Emam2018Inf}, we measured accuracy in terms
of \emph{recall} and \emph{specificity} of the inferred models. Recall measures the
ability of the inferred models to accept ``positive'' logs that correspond to
correct or valid behaviors of the system. In contrast, specificity
measures the ability of the inferred models to reject ``negative'' logs that
correspond to incorrect or invalid behaviors.

We computed recall and specificity by using $k$-folds
cross-validation ($k=10$) with the synthesis of negative logs, which has also been
used in previous work~\cite{walkinshaw2016inferring, mariani2017gk, Emam2018Inf}
in the area of model inference. This technique randomly partitions a set of
positive logs into $k$ non-overlapping folds: $k-1$ folds are used as input of
the model inference tool, while the remaining fold is used as ``test set'', to
check whether a model correctly accepts positive logs
in the test set. The procedure is repeated $k$ times until all folds
have been considered exactly once as the test set.
Further, to check whether the inferred model correctly rejects
negative logs, the technique synthesizes negative logs from the positive logs in
the test set by
\begin{inparaenum}[(1)]
	\item randomly adding a log entry or
	\item randomly swapping multiple log entries.
\end{inparaenum}
To make sure the synthesized logs really capture invalid system
behaviors, we checked whether the sequence of transactional events of
each synthesized negative log could not be produced by the publicly
available model of the system (for the non-proprietary systems) or did
not appear in the (sub)sequences of transactional events of the
positive logs (for the proprietary systems).

For each fold, if the inferred model successfully accepts a positive log in
the test set, the positive log is classified as True Positive (TP); otherwise,
the positive log is classified as False Negative (FN). Similarly, if an inferred
model successfully rejects a negative log in the test set, the negative log is
classified as True Negative (TN); otherwise, the negative log is classified as
False Positive (FP). Based on the classification results, we computed the recall
and specificity using the same formulae shown in \S~\ref{sec:RQ1:setup}.

Note that $k$-folds cross-validation measures how accurate model inference
is for a given set of logs. In other words, it
measures accuracy based on the given logs of a system, not based on the
``correct'' model that represents the functional behavior of the system. Thus,
even if the logs are inaccurately pre-processed by \app, a model inferred from a
set of logs could achieve 1 in both recall and specificity. Nevertheless, we
used $k$-folds cross-validation since there is
\begin{inparaenum}[(1)]
	\item no standard way to measure similarity between two FSMs (to
measure the accuracy of an inferred FSM with respect to a correct FSM) and
	\item no correct model for the proprietary systems.
\end{inparaenum}

To account for randomness in the generation of $L_\mathit{org}$ and the
injection of operational messages, we repeated the 10-folds cross-validation 30
times for each system in our benchmark and applied the non-parametric Wilcoxon
signed-ranks test to assess the statistical significance of the difference in
accuracy between the models inferred from $L_\mathit{org}$ and those
inferred from $L_\mathit{cl}$.

\subsubsection{Results}
Table~\ref{table:RQ3} shows the results of the impact of \app on model
inference accuracy, grouped by NR and system.
Under the \emph{Recall} column, sub-column \emph{$L_x$} indicates the recall on
$L_x$ where $x \in \{\mathit{org}, \mathit{cl}\}$, and sub-column
\emph{$\Delta_R$} indicates the difference in recall between $L_\mathit{org}$ and
$L_\mathit{cl}$ in percentage points ($pp$). The sub-columns under the
\emph{Specificity} column follow the same structure,
with sub-column \emph{$\Delta_S$} indicating the difference in specificity
between $L_\mathit{org}$ and $L_\mathit{cl}$.
All the difference values not marked with asterisk  are statistically
significant ($p\text{-value} < 0.01$).

\begin{table}
\footnotesize
\caption{Impact of \app on Model Inference Accuracy}\label{table:RQ3}
\begin{tabular}{llrrrrrr}
\toprule
 & & \multicolumn{3}{c}{Recall} & \multicolumn{3}{c}{Specificity}  \\
\cmidrule(r){3-5}\cmidrule(r){6-8}
NR & System     & $L_\mathit{org}$   & $L_\mathit{cl}$ & $\Delta_R$ ($pp$) & $L_\mathit{org}$   & $L_\mathit{cl}$ & $\Delta_S$ ($pp$) \\
\midrule
\multirow{2}{*}{-}    & \texttt{SYS1}            & 0.060 & 0.663 & 60.3  & 1.000 & 1.000 & 0.0*    \\
                      & \texttt{SYS2}            & 0.967 & 0.967 & 0.0   & 0.775 & 1.000 & 22.5  \\
\midrule
\multirow{11}{*}{0.3} & \texttt{CVS}             & 0.070 & 0.956 & 88.6  & 0.984 & 1.000 & 1.6   \\
                      & \texttt{Lucane}          & 0.347 & 0.974 & 62.7  & 0.825 & 1.000 & 17.5  \\
                      & \texttt{RapidMiner}      & 0.086 & 0.943 & 85.7  & 0.985 & 0.891 & -9.4  \\
                      & \texttt{SSH}             & 0.206 & 0.514 & 30.7  & 0.877 & 0.977 & 10.0  \\
                      & \texttt{DatagramSocket}  & 0.000 & 0.903 & 90.3  & 1.000 & 0.997 & -0.3* \\
                      & \texttt{MultiCastSocket} & 0.058 & 0.998 & 94.0  & 0.990 & 0.988 & -0.3* \\
                      & \texttt{Socket}          & 0.006 & 0.971 & 96.5  & 1.000 & 0.999 & -0.1* \\
                      & \texttt{Formatter}       & 0.466 & 0.855 & 38.9  & 0.671 & 0.947 & 27.6  \\
                      & \texttt{StringTokenizer} & 0.440 & 0.890 & 45.0  & 0.667 & 0.889 & 22.2  \\
                      & \texttt{TCPIP}           & 0.179 & 0.842 & 66.2  & 0.898 & 0.878 & -2.0* \\
                      & \texttt{URL}             & 0.001 & 1.000 & 99.9  & 1.000 & 1.000 & 0.0*  \\
\midrule
\multirow{11}{*}{0.7} & \texttt{CVS}             & 0.296 & 0.975 & 67.9  & 0.800 & 1.000 & 20.0  \\
                      & \texttt{Lucane}          & 0.286 & 0.981 & 69.6  & 0.763 & 1.000 & 23.7  \\
                      & \texttt{RapidMiner}      & 0.503 & 0.940 & 43.7  & 0.796 & 0.900 & 10.4  \\
                      & \texttt{SSH}             & 0.552 & 0.532 & -2.0* & 0.518 & 0.960 & 44.2  \\
                      & \texttt{DatagramSocket}  & 0.224 & 0.049 & -17.5 & 0.828 & 1.000 & 17.2  \\
                      & \texttt{MultiCastSocket} & 0.464 & 0.501 & 3.7*  & 0.586 & 1.000 & 41.4  \\
                      & \texttt{Socket}          & 0.223 & 0.031 & -19.1 & 0.827 & 1.000 & 17.3  \\
                      & \texttt{Formatter}       & 0.538 & 0.849 & 31.1  & 0.535 & 0.848 & 31.3  \\
                      & \texttt{StringTokenizer} & 0.556 & 0.838 & 28.3  & 0.509 & 0.726 & 21.7  \\
                      & \texttt{TCPIP}           & 0.529 & 0.754 & 22.5  & 0.557 & 0.788 & 23.1  \\
                      & \texttt{URL}             & 0.425 & 0.062 & -36.4 & 0.801 & 0.999 & 19.8  \\
\midrule
\multicolumn{2}{l}{Average for all cases} & 0.312 & 0.750 & 43.8  & 0.800 & 0.949 & 15.0 \\
\multicolumn{2}{l}{Standard deviation}    & 0.240 & 0.311 & 40.4  & 0.169 & 0.077 & 14.1 \\
\bottomrule
\end{tabular}
\end{table}

On average, $\Delta_R$ is 43.8\thinspace $pp$ (with a standard deviation of 40.4)
and $\Delta_S$ is 15.0\thinspace $pp$ (with a standard deviation of 14.1).
These results mean that pre-processing the input
logs using \app may significantly increase the accuracy of model
inference. The intuitive explanation is that by removing randomly
interleaved (operational) messages, \app reduces the noise in the
input logs, increasing the likelihood of inferring a more accurate model.

However, there are some cases where $\Delta_R$ is significantly
negative, such as (for NR = 0.7), \texttt{DatagramSocket}
(-17.5\thinspace $pp$), \texttt{Socket} (-19.1\thinspace $pp$), and
\texttt{URL} (-36.4\thinspace $pp$). These cases are mainly due to low
recall values on $L_\mathit{cl}$ (0.049 for \texttt{DatagramSocket},
0.031 for \texttt{Socket}, and 0.062 for \texttt{URL}). Considering
the perfect accuracy of \app when NR = 0.7, such exceptionally low
recall values indicate that inferring an accurate model from low-quality logs
is challenging, even when the logs do not contain any noise.

We remark that, for the same \texttt{Type-B} systems, recall on
$L_\mathit{cl}$ when NR = 0.3 is much higher than when NR = 0.7.
This means that in low-quality logs (as discussed
in \S~\ref{sec:RQ1:results}), the incorrect removal by \app of some
``low-quality'' transactional messages (which are randomly
interleaved) contributes to improving the quality of the logs, which
leads to infer more accurate models.

In terms of specificity, there is only one case (\texttt{RapidMiner}
with NR = 0.3) for which $\Delta_S$ is significantly negative
(-9.4\thinspace $pp$); nevertheless, the same case has a large
$\Delta_R$ of 85.7\thinspace $pp$. This means that pre-processing the
input logs using \app decreases specificity but it does so to a
limited extent while greatly increasing recall, resulting in a clearly
more accurate model.

To summarize, the answer to RQ3 is that pre-processing the input logs
of model inference using \app significantly improves the accuracy of
the inferred models by increasing their ability to accept correct system
behaviors (+43.8\thinspace $pp$ on average with a standard deviation
of 40.4) and to reject incorrect system behaviors (+15.0\thinspace $pp$
on average with a standard deviation of 14.1).
However, we also found that the accuracy of inferred models could be
very low (recall below 0.1) in some cases when low-quality logs are
used as input, even if the logs are accurately pre-processed by
\app. This implies that, in practice, it is important to ensure that
the granularity of logging statements is consistent with the
functional behavior of the system being logged, as discussed in
\S~\ref{sec:RQ1:results}.

\subsection{Threats to Validity}\label{sec:discussion}
As noted in Section~\ref{sec:preliminary}, different log parsing
techniques may lead to different sets of templates for the same logs,
affecting the accuracy of \app. For example, if a template is over-generalized,
meaning log entries with different events are regarded as log entries
of the same event template, then the dependency score of the template
is likely to decrease as the number of log entries of the template
increases. To mitigate such threats related to log entry
templates, we did not make any modifications on the events given in
the non-proprietary models. For the proprietary systems, we used a
state-of-the-art log parsing technique~\cite{messaoudi2018search} to
extract templates from the unstructured logs and then improved a few
under/over-generalized templates with the help of a domain expert.

Using a specific model inference tool (MINT) is another
potential factor that may affect our results. However, we expect that applying other model
inference techniques would not change the trends in results since the
fundamental principles at the basis of model inference are very similar.
Nevertheless, an experimental comparison across alternative model
inference tools is left for future work.

\section{Related Work}\label{sec:related-work}
The closest work to \app is the automated operational message
filtering technique by \citet{8030620}, proposed as the pre-processing
step of their anomaly detection approach. However, no implementation or 
even a precise algorithm was provided by the authors, as previously 
noted in \S~\ref{sec:RQ1:setup}. According to the textual description 
of the technique, its underlying idea is similar to the dependency analysis of \app. 
The technique is based on how frequently two templates $x$ and $y$ 
co-occur within a small distance threshold $h_1$ in logs, without
distinguishing the dependency for which $x$ could be a \emph{cause} of
$y$ and the dependency for which $x$ could be a \emph{consequence} of
$y$. Thus, two operational templates could have a high dependency
score if they are frequently interleaved within $h_1$ in random order.
On the other hand, \app considers the two dependencies using separate
\emph{forward} and \emph{backward} dependency scores.  Furthermore,
this technique  requires another threshold $h_2$
to distinguish operational templates and transactional templates,
whereas \app automatically distinguish them using a clustering-based
heuristic (detailed in \S~\ref{sec:clustering}).  As a result,
the performance of this filtering technique is
largely dependent on $h_1$ and $h_2$, which have to be determined by
an engineer and for whose estimation the article does not provide any guideline.
On the contrary, \app only requires to specify the
periodicity threshold for the periodicity analysis, which can be
easily determined from the timestamp granularity of the log entries.

In the area of (business) process mining, there have been proposals
for techniques for noise filtering using a specific temporal
relationship between events~\cite{7579568, 8818438} and for live event
streams~\cite{Zelst2018Filtering}. However, since they mainly target
outliers or infrequent behaviors, they cannot be applied for removing
operational messages.

\section{Conclusion}
\label{sec:conclusion}
In this paper, we propose \app, a novel technique for effectively
removing operational messages---recording the operational state of the
system---from logs, and keeping only transactional messages, which
record the functional behavior of the system. \app distinguishes
between operational and transactional messages by analyzing their
periodicity and their dependencies (in terms of co-occurrence). The
logs pre-processed with \app can then be used for several software
engineering tasks, such as model inference, which is the focus of
this paper.

We evaluated \app on two proprietary and 11 publicly available log
datasets. The experimental results show that \app, on average, can
remove 98\% of the operational messages and preserve 81\% of the
transactional messages.
Furthermore, using logs pre-processed with \app decreases the
execution time of model inference (with a speed-up ranging from 1.5 to
946.7 depending on the characteristics of the system) and improves the
accuracy of the inferred models, by increasing their ability to accept 
correct system behaviors (+43.8\thinspace $pp$ on average) and to reject 
incorrect system behaviors (+15.0\thinspace $pp$ on average).
In all cases, the performance and impact of \app depend on the quality
of the input logs. This implies that, in practice, it is important to
ensure that the granularity of logging statements is consistent with
the functional behavior of the system being logged.

As part of future work, we plan to study the impact of \app on
different model inference tools~\cite{Beschastnikh2011Lev,
  mariani2017gk}, and to investigate in more detail the relationship
between the quality of logs and the performance of \app.

\bibliographystyle{ACM-Reference-Format}
\bibliography{log-cleaner}

\end{document}